\documentstyle[preprint,aps]{revtex}
\begin{document}
\preprint{DAMTP/R-95/8}
\draft
\tighten
\title{\large\bf Duality between Electric
and Magnetic Black Holes }
\author{S.W. Hawking\footnote{E-mail: swh1@damtp.cam.ac.uk} and
Simon F. Ross\footnote{E-mail: S.F.Ross@damtp.cam.ac.uk}}
\address{Department of Applied Mathematics and Theoretical Physics \\
University of Cambridge, Silver St., Cambridge CB3 9EW}
\date{\today}
\maketitle
\begin{abstract}
A number of attempts have recently been made to extend the conjectured
$S$ duality of Yang Mills theory to gravity.  Central to these
speculations has been the belief that electrically and magnetically
charged black holes, the solitons of quantum gravity, have identical
quantum properties.  This is not obvious, because although duality is
a symmetry of the classical equations of motion, it changes the sign
of the Maxwell action.  Nevertheless, we show that the chemical
potential and charge projection that one has to introduce for electric
but not magnetic black holes exactly compensate for the difference in
action in the semi-classical approximation.  In particular, we show
that the pair production of electric black holes is not a runaway
process, as one might think if one just went by the action of the
relevant instanton.  We also comment on the definition of the entropy
in cosmological situations, and show that we need to be more careful
when defining the entropy than we are in an asymptotically-flat case.
\end{abstract}
\pacs{PACS numbers:04.70.Dy, 04.60.-m, 04.40.Nr}
\narrowtext

\section{Introduction}
\label{intro}

The idea of duality has received considerable attention recently,
particularly in the context of string theory. This is a subject with a
long history, which may be traced back to Olive and Montonen's
conjectured duality in Yang-Mills theory \cite{ovmon}. In the $N=4$
Yang-Mills theory, one has two kinds of particle: small fluctuations
in the scalar or Yang-Mills fields, and magnetic monopoles. The small
fluctuations couple to the Yang-Mills field like electrically charged
particles couple to the Maxwell field. They are therefore regarded as
electrically charged elementary states. But the magnetic monopoles,
which are the solitons of the theory, can also claim to be regarded as
particles.  Olive and Montonen conjectured \cite{ovmon} that there was
a dual Yang-Mills theory, with coupling constant $g'=1/g$. Monopoles
in the dual theory would behave like the elementary electrically
charged states of the original theory, and vice versa.  This concept
of duality was later extended to a lattice of theories related by the
discrete group $SL(2,Z)$. There is some evidence that the low energy
scattering of monopoles is consistent with what one would expect from
this duality, which is called $S$-duality, but no proof has been given
that it goes beyond a symmetry of the equations of motion to a
symmetry of the full quantum theory.

Despite this lack of proof, there has been extensive speculation on
how $S$-duality could extend to gravity and string-inspired
supergravity theories \cite{sen}.  The suggestion is that extreme,
non-rotating black holes should be identified as the solitons of the
theory. These states do have some particle-like properties, as there
are families of electric and magnetic black holes, which fall into
multiplets under the action of the global supersymmetry group at
infinity.  The similarity with other solitons has been increased by
our recent discovery \cite{entar} that all extreme black holes have
zero entropy, as one would expect for elementary particles.  However,
the original Montonen and Olive idea of duality was supposed to relate
electrically charged elementary states, or small fluctuations in the
fields, with magnetically charged monopoles, or solitons. But in the
gravitational case there are both magnetically and electrically
charged solitons. This has led people to try to identify extreme black
holes, the solitons, with electrically or magnetically charged
elementary states in string theory \cite{rususs,duff}. The only
evidence so far is that one can find black holes with the same masses
and charges as a certain class of elementary states \cite{duff}. But
this is not very surprising, because the masses are determined by the
charges and Bogomol'nyi bounds in both cases.

Behind all these attempts to extend $S$-duality to extreme black holes
is the idea that electrically and magnetically charged black holes
behave in a similar way. This is true in the classical theory, because
duality between electric and magnetic fields is a symmetry of the
equations. This does not, however, imply that it is a symmetry of the
quantum theory, as the action is not invariant under duality. The
Maxwell action is $F^2 = B^2-E^2$, and it therefore changes sign when
magnetic fields are replaced by electric. The purpose of this paper is
to show that despite this difference in the action, the semi-classical
approximations to the Euclidean path integral for dual electric and
magnetic solutions are identical, at least where we have been able to
evaluate them. In particular, we show that the rate at which black
holes are pair created in  cosmological and electromagnetic backgrounds is
duality-invariant.

We will now define our terms more precisely.  It is well known that
the Einstein-Maxwell equations exhibit duality. One can replace
magnetic fields with electric fields and a solution remains a
solution. More precisely, if $(g,F)$ are a metric and field tensor
that satisfy the field equations, then $(g,*F)$ also satisfy the
equations, where $*F$ is the Lorentzian dual of $F$, that is,
\begin{equation} \label{Ldual}
*F_{\mu\nu} = \frac{1}{2} \epsilon_{\mu\nu\rho\sigma} F^{\rho\sigma},
\end{equation}
with $\epsilon_{0123} = \sqrt{-g}$, and $g$ the determinant of the
metric. If $F$ represents a magnetic field, $*F$ will represent an
electric field, referred to as the dual electric field. In particular,
for every magnetically charged black hole solution, there is a
corresponding electrically charged black hole solution. This
electric-magnetic duality extends to theories with a dilaton. The only
difference is that one now takes
\begin{equation}
*F_{\mu\nu} = \frac{1}{2} e^{-2a\phi} \epsilon_{\mu\nu\rho\sigma}
F^{\rho\sigma}
\end{equation}
and
\begin{equation}
\phi \to -\phi,
\end{equation}
where $\phi$ is the dilaton field. We will, however, restrict
attention to the duality (\ref{Ldual}) in Einstein-Maxwell theory for
the sake of simplicity.

Now, if $g_{\mu\nu}$ is a Lorentzian metric, its determinant will be
negative, so $\epsilon_{\mu\nu\rho\sigma}$ as defined above will be
real. However, if $g_{\mu\nu}$ is a Euclidean metric, its determinant
will be positive, and thus $\epsilon_{\mu\nu\rho\sigma}$ will be
imaginary. That is, the Lorentzian duality (\ref{Ldual}) takes real
magnetic fields to real electric fields in Lorentzian space, but real
magnetic fields to imaginary electric fields in Euclidean space. This
is consistent, as an electric field that is real in a Lorentzian space
is imaginary in its Euclidean continuation. One might therefore think that, in
using the Euclidean path integral, one should use Euclidean duality
instead of Lorentzian duality, and replace magnetic fields with
electric fields that were real in Euclidean space. That is, perhaps
one should take
\begin{equation} \label{Edual}
*F_{\mu\nu} =  \frac{i}{2} \epsilon_{\mu\nu\rho\sigma} F^{\rho\sigma}
\end{equation}
instead of (\ref{Ldual}). This duality also has the advantage that it
leaves the Maxwell action unchanged. However, it
reverses the sign of the energy momentum tensor, so the solutions
would have different geometry. That is, if $*F$ is given by
(\ref{Edual}), then it is no longer true that $(g,*F)$ satisfy the
field equations whenever $(g,F)$ do. In particular, there is no
extreme black hole solution with real electric fields in Euclidean
space. It seems therefore that if duality is to be a symmetry of black
holes, it must be a duality between real electric and magnetic fields
in Lorentzian space, rather than in Euclidean space.

There is then a difference in action between the dual electric and
magnetic solutions. What effect will this have? One of the most
interesting applications of the Euclidean path integral approach is
the study of semi-classical instabilities, or tunnelling processes.
One uses instantons, Euclidean solutions of the field equations, to
estimate the rate at which such classically-forbidden tunnelling
processes occur. The rate at which a process occurs is just given by
the partition function $Z$, defined by
\begin{equation} \label{parf}
Z = \int d[g] d[A] e^{-I},
\end{equation}
where the integral is subject to some appropriate boundary conditions
at infinity. When there is a Euclidean solution which satisfies the
boundary conditions, we can approximate the integral by the
saddle-point, which gives $Z \approx e^{-I}$, where $I$ is the action of
the instanton, so it would seem that the difference in action between
dual solutions must surely imply a difference in the rate for such
processes. In particular, the Euclidean black hole solutions can be
used as instantons for black hole nucleation or pair creation, and we
might therefore think that electrically and magnetically charged black
holes should be produced at different rates. However, a more careful
analysis of the partition function shows that this is not the case.

The point is that magnetic and electric solutions differ not only in
their actions, but in the nature of the boundary conditions we can
impose on them. If we consider a single black hole, we can choose a
particular charge sector in the magnetic case, but we have to introduce
a chemical potential for the charge in the electric case. That is to
say, we can impose the magnetic charge as a boundary condition at
infinity, but we can only impose the chemical potential, and not the
electric charge, as a boundary condition in the electric case. Thus
the partition function in the magnetic case is a function of the
temperature and charge, $Z(\beta,Q)$, while in the electric case
the partition function is a function of the chemical potential
$\omega$, rather than $Q$, $Z(\beta,\omega)$. It is not surprising to
find that these two quantities differ. What we need to do is
obtain a partition function $Z(\beta,Q)$ in the electric case. To do
this, we must introduce a charge projection operator \cite{quhair}.

The introduction of the charge projection operator is like
performing a Fourier transform on the wavefunction, to trade $\omega$
for its canonically conjugate momentum $Q$. The effect of this
transform is to make the partition function as a function of charge
the same for the electrically and magnetically charged black
holes. The difference in action precisely cancels the additional term
introduced in the partition function by the Fourier transform.

We can also calculate $Z(\beta,Q)$ in the electric case directly, by
using (\ref{parf}) with an action which is adapted to holding the
electric charge fixed. To make the action give the classical equations
of motion under a variation which holds the electric charge on the
boundary fixed, we need to include an additional surface term in the
action. This will make the action of dual electric and magnetic
solutions identical.

We are particularly interested in instantons describing black hole
pair creation. To obtain pair creation of black holes, one has to
have some force that is pulling the holes apart. The case that has
been extensively studied is the formation of charged black holes in a
background electric or magnetic field
\cite{entar,garstrom,dgkt,dggh,2u1}. Here the negative electromagnetic
potential energy of the holes in the background electric or magnetic
field can compensate for the positive rest mass energy of the black
holes. The pair creation of magnetically-charged black holes in a
background magnetic field has been the subject of most work in this
area, and the action and pair creation rate for this case have been
calculated in \cite{garstrom,dgkt}. It was assumed in earlier work
that the treatment of the electric case was a trivial extension of the
magnetic; we now realize that this is not in fact the case. We
consider the pair creation of electric black holes in a background
electric field, and show by calculating $Z(\beta,Q)$ directly that the
pair creation rate in this case is the same as in the magnetic case.

The effective cosmological constant in the inflationary period of the
universe can also accelerate objects away from each other, and so it
should be possible to find instantons describing the pair production
of black holes in a cosmological background. In the case without gauge
fields, the relevant solution is the Schwarzschild de Sitter
metric. This has been interpreted in the past as a single black hole
in a de Sitter universe, but it really represents a pair of black
holes at antipodal points on the three sphere space section of the de
Sitter universe, accelerating away from each other. If one takes $t
=i\tau$, one obtains a Euclidean metric. One can remove the conical
singularities in this metric if the black hole and cosmological
horizons have the same temperature. For the Schwarzschild de Sitter
metric, this occurs in the limiting case known as the Nariai metric,
which is just the analytical continuation of $S^2\times S^2$, with
both spheres having the same radius \cite{desit}.

If you cut this solution in half, you get the amplitude to propagate
from nothing to a three surface $\Sigma$ with topology $S^2 \times
S^1$ according to the no boundary proposal.  One can regard $S^2
\times S^1$ as corresponding to the space section of the Nariai
universe, which will settle down to two black holes in de Sitter space
(see \cite{desit} for more details). The action of $S^2 \times S^2$ is
$I=- 2 \pi /\Lambda$. This is greater than the action $I= - 3 \pi/
\Lambda$ of $S^4$, which corresponds to de Sitter space. Thus the
amplitude to pair create neutral black holes in de Sitter space is
suppressed, as one would hope.  \footnote{If one were to use the
tunnelling proposal \cite{vil,lin} instead of the no boundary
proposal, one would find that the probability of the pair creation of
neutral black holes was enhanced rather than suppressed relative to
the probability for the spontaneous formation of a de Sitter universe.
This is further evidence against the tunnelling proposal.}

One can also consider the pair creation of electrically or
magnetically charged black holes in de Sitter space. Here the relevant
solutions are the Reissner-Nordstr\"om de Sitter metrics, which can
again be extended to Euclidean metrics. More than one instanton can be
constructed in this case; these instantons are discussed in more
detail in \cite{moss,romans,robb}. We will consider only the simplest
case, where the instanton is again $S^2 \times S^2$, but where the
spheres now have different radii. The action for the magnetic
instanton is less negative than the neutral case. Thus the pair
creation of magnetic black holes is suppressed relative to neutral
black holes, which is itself suppressed relative to the background de
Sitter space. All this is what one might expect on physical
grounds. But in the electric case, the action is less than the action
of the neutral case, and can be less than the action of the background
de Sitter space if the electric charge is large enough. This at first
seemed to suggest that de Sitter space would be unstable to decay by
pair production of electrically-charged black holes.

Presumably, we have to apply a charge projection operator to obtain
comparable partition functions here, as in the single black hole
case. However, the $S^2 \times S^2$ instanton has no boundary, so we
at first thought that it wasn't possible to have a chemical potential
in this case. However, as we said above, what we actually want to
consider is the amplitude to propagate from nothing to a three surface
$\Sigma$ with topology $S^2 \times S^1$, and we can impose the
potential on the boundary $\Sigma$. The instanton giving the
semi-classical approximation to this amplitude is just half of $S^2
\times S^2$. In the magnetic case, the magnetic charge can be given as
a boundary condition on this surface, but in the electric case, the
boundary gives only the potential $\omega$. If we again make the
Fourier transform to trade $\omega$ for $Q$, the semi-classical
approximation to the wavefunction as a function of charge is the same
for the electrically and magnetically charged black holes. Thus the
pair creation of both magnetic and electric black holes is suppressed
in the early universe.

We will also discuss the entropy of black hole solutions. For the
asymptotically-flat black holes, the partition function $Z(\beta,Q)$
can be interpreted as the canonical partition function, while
$Z(\beta,\omega)$ can be interpreted as the grand canonical partition
function. Using the instantons to approximate the partition function,
we can show that the entropy of the asymptotically-flat black holes is
$S = {\cal A}_{bh}/4$ for both electrically and magnetically charged
black holes.

For the cosmological solutions, the square of the wavefunction
$\Psi(Q,\pi^{ij}=0)$ can be regarded as the density of states or
microcanonical partition function. Thus the entropy is just given by
the ln of the wavefunction. Using the instantons to approximate this
density of states, we find that the entropy is $S = {\cal A}/4$, where
${\cal A}$ is the total area of all the horizons in the instanton.

In section \ref{chproj}, we review the calculation of the action for
the Reissner-Nordstr\"om black holes, and the introduction of the
charge projection operator. In section \ref{desitsec}, we describe the
Reissner-Nordstr\"om de Sitter solution, and derive an instanton which
can be interpreted as describing black hole pair production in a
background de Sitter space. We then calculate its action. We go on to
argue, in section \ref{cosm}, that a charge projection can be
performed in this case as well, and that the partition function as a
function of the charge is the same in the electric and magnetic cases.
In section \ref{elecesec}, we review the electric Ernst solution, and
obtain the instanton which describes pair creation of
electrically-charged black holes in an electric field. In section
\ref{elacsec}, we calculate the action for this instanton, and thus
obtain the pair creation rate. In section \ref{entsec}, we review the
derivation of the entropy for the Reissner-Nordstr\"om black holes,
and discuss its definition for the Reissner-Nordstr\"om de Sitter
solutions.

\section{Action and charge projection in Reissner-Nordstr\"om}
\label{chproj}

Let us first consider asymptotically-flat black hole solutions. To
simplify the later calculation of the entropy, we will evaluate the
action of these black holes by a Hamiltonian decomposition, following
the treatment given in \cite{haho}. If there is a Maxwell or
Yang-Mills field, one takes spatial components of the vector potential
$A_i$ as the canonical coordinates on three-surfaces of constant
time. The conjugate momenta are the electric field components
$E^i$. The time component $A_t$ of the potential is regarded as a
Lagrange multiplier for the Gauss law constraint ${\rm div} E=0$. Let
us first assume that the manifold has topology $\Sigma \times
S^1$. Then the action is
\begin{equation} \label{act}
I= -\int dt \left[\int_{\Sigma_t}
  (p^{\mu\nu}\   {}^3 \dot g_{\mu\nu} + E^i \dot{A}_i) - H \right].
\end{equation}
There is a well-known ambiguity in the gravitational action for
manifolds with boundary, as one can add any function of the boundary
data to the action, and its variation will still give the same
equations of motion \cite{brown}. We will adopt the approach of
\cite{haho}, and require that the action of some suitable background
vanish. We define a suitable background to be one which agrees with
the solution asymptotically, that is, which induces the same metric
and gauge fields on $S^2_\infty$. If we assume that the background is a
solution of the equations of motion, the Hamiltonian $H$ is \cite{bmy}
\begin{eqnarray} \label{ham}
 H &=& \frac{1}{8 \pi} \int_{\Sigma_t} (N{\cal H} +N^i {\cal H}_i + N A_t
{\rm div} E) \\  &&-
{1\over 8\pi}
   \int_{S^2_\infty} [N (^2 K - ^2 K_0) + N^i p_{ij} + 2 N A_t(E -
E_0)], \nonumber
\end{eqnarray}
where $^2 K$ is the extrinsic curvature of the boundary $S^2_{\infty}$
of the surface $\Sigma_t$, $E$ is the electric field, and $^2 K_0$ and
$E_0$ represent these quantities evaluated in the background.

In order to get the action in this canonical form, we have had to
integrate by parts the terms in the action involving spatial gradients
of $A_t$. This produces the $A_t$ surface term in the
Hamiltonian. This surface term is zero for magnetic monopoles and
magnetic black holes. It is also zero for any solution with electric
fields, but no horizons, because one can choose a gauge in which $A_t$
vanishes at infinity. Thus, the existence of this surface term in the
Hamiltonian does not seem to have been generally noticed. However, it
is non-zero for electrically charged black holes, because the gauge
transformation required to make $A_t=0$ at infinity is not regular on
the horizon.

One can pass from a Lorentzian black hole solution to a Euclidean one
by introducing an imaginary time coordinate $\tau = -i t$. One then has
to identify $\tau$ with period $\beta = 2 \pi/ \kappa$ to make the
metric regular on the horizon, where $\kappa$ is the surface gravity
of the horizon.  One can then use the relation between the action and
the Hamiltonian to calculate the action of the Euclidean black hole
solution. As the solution is static, the Euclidean action (\ref{act})
is $\beta$ times the Hamiltonian. However, the Euclidean section for a
non-extreme black hole does not have topology $\Sigma \times S^1$, and
so (\ref{act}) only gives the action of the region swept out by the
surfaces of constant $\tau$. This is the whole of the Euclidean
solution, except for the fixed point locus of the time translation
killing vector on the horizon. The contribution to the action from the
corner between two surfaces $\tau_1$ and $\tau_2$ is
\begin{equation}
\frac{\kappa}{8\pi} (\tau_2 - \tau_1) {\cal A}_{bh},
\end{equation}
where ${\cal A}_{bh}$ is the area of the horizon. Thus the
action is $I= \beta H -{\cal A}_{bh}/4$ \cite{ecs}.  For solutions of
the field equations, the three-surface integral vanishes, because of
the gravitational and electromagnetic constraint equations. Thus, the
value of the Hamiltonian comes entirely from the surface terms.

Now we will calculate the action in this way for the nonextreme
electric and magnetic Reissner-Nordstr\"om solutions.  Recall that the
Reissner-Nordstr\"om metric is given by
\begin{eqnarray} \label{exrn}
ds^2  &=& -\left( 1-\frac{2M}{ r} + \frac{Q^2}{ r^2}\right) dt^2
+ \left( 1-\frac{2M}{ r} +\frac{Q^2}{ r^2}\right)^{-1} dr^2  \nonumber
\\ &&+r^2( d\theta^2 + \sin^2 \theta d \phi^2),
\end{eqnarray}
where $M$ is the mass and $Q$ is the charge of the black hole. The
gauge potential for this solution is
\begin{equation} \label{mag}
F = Q \sin \theta d\theta \wedge d\phi
\end{equation}
for a magnetically-charged solution, and
\begin{equation} \label{elec}
F = - \frac{Q}{r^2} dt \wedge dr
\end{equation}
for an electrically-charged solution. We will not consider dyonic
solutions.  The metric has two horizons, at $r=r_\pm = M \pm \sqrt{
M^2 -Q^2}$. We analytically continue $t \to i \tau$, and identify
$\tau$ with period $\beta = 2 \pi / \kappa$, where $\kappa = (r_+ -
r_-)/2 r_+^2 $ is the surface gravity of the horizon at $r=r_+$. The
surfaces of constant $\tau$ meet at the event horizon $r=r_+$, whose
area is
\begin{equation} \label{arearn}
{\cal A}_{bh}  = 4\pi r_+^2 = \frac{4\pi}{\kappa} (M -Q U),
\end{equation}
where $U = Q/r_+$. The second equality
is obtained by exploiting the definitions of $r_\pm$ and $\kappa$.

If we consider the magnetically charged black hole solution, the gauge
potential will be
\begin{equation} \label{magrn}
A  = Q (1 -\cos \theta)d\phi,
\end{equation}
where we have chosen a gauge which is regular on the axis $\theta =0$.
For a magnetic black hole, the electromagnetic surface term in the
Hamiltonian vanishes, and the Hamiltonian is just given by the
gravitational surface term. However, as the background spacetime
usually used to calculate the Hamiltonian for the Reissner-Nordstr\"om
black holes is just periodically-identified flat space, this surface
term is equal to the usual ADM mass \cite{haho}. Thus the Hamiltonian
is simply
\begin{equation}
H=M,
\end{equation}
and if $\tau$ is identified with period $\beta= 2 \pi / \kappa$, the
action is
\begin{equation} \label{magac}
I = \beta M - {\cal A}_{bh}/4 = \frac{\pi}{\kappa} ( M + Q U).
\end{equation}

For the electrically charged black hole solution, the gauge potential
is
\begin{equation} \label{elrn}
A = -i(Q/r - \Phi)d\tau,
\end{equation}
where $\Phi = U$ is the potential at infinity
and we have chosen a gauge which is regular on the black hole
horizon. Note that this gauge potential is pure imaginary, as we have
analytically continued $t \to i\tau$. We take the point of view that
one should simply accept that the gauge potential in Euclidean space
is imaginary; if one analytically continued the charge to obtain a
real gauge potential, the metric would be changed, and one could no
longer sensibly compare the electric and magnetic solutions, as they
would no longer be dual solutions. In this case, the Hamiltonian is
still just equal to the surface term, but now the electromagnetic
surface term survives as well. The Hamiltonian can now be calculated
to be
\begin{equation}
H = M - Q \Phi,
\end{equation}
and we see that $\Phi$ may be interpreted as the electrostatic
potential in this case. Thus, if $\tau$ is identified with period
$\beta = 2 \pi / \kappa$, the action is
\begin{equation} \label{elac}
I = \beta(M - Q \Phi) - {\cal A}_{bh}/4 = \frac{\pi}{\kappa} (M - Q U),
\end{equation}
as asserted in \cite{actionint}. If we were to calculate the action
directly, as was done in \cite{actionint}, we would find that the sign
difference of the $Q U$ term in the action is due to the fact that
$F^2 = 2Q^2/r^4$ for the magnetic solution, but $F^2 = -2Q^2/r^4$ for
the electric solution.

As we have said in the introduction, the na\"{\i}ve expectation that the
rate of pair creation is simply approximated by the action ignores an
important difference between the electric and magnetic cases. The
partition function is
\begin{equation} \label{pathin}
Z = \int d[g] d[A] e^{-I[g,A]},
\end{equation}
where the integral is over all metrics and potentials inside a
boundary $\Sigma^\infty$ at infinity, which agree with the given
boundary data on $\Sigma^\infty$. Now for the Euclidean black holes,
the appropriate boundary is $\Sigma^\infty = S^2_\infty \times S^1$,
and the boundary data are the three-metric $h_{ij}$ and gauge
potential $A_i$ on the boundary at infinity. In the magnetic case, one
can evaluate the magnetic charge by taking the integral of $F_{ij}$
over the $(\theta,\phi)$ two-sphere lying in the boundary, so the
magnetic charge is a boundary condition. That is, we are evaluating
the partition function in a definite charge sector. In the electric
case, however, $A_i$ is constant on the boundary, so all we can
construct is an integral of it over the boundary. This is the chemical
potential $\omega = \int A_\tau d\tau$, where we define this integral
to be in the direction of increasing $\tau$. That is, we are
evaluating the partition function in a sector of fixed $\omega$.  This
can be written in a shorthand form as $Z(\beta,\omega)$. To obtain the
partition function in a sector of definite charge, we have to
introduce a charge projection operator in the path integral
\cite{quhair}. This gives\footnote{There is a sign difference between
this expression and the analogous expression in \cite{quhair}, but
this is just due to a difference of conventions.}
\begin{equation} \label{proj1}
Z(\beta,Q)  = \frac{1}{2\pi} \int_{-\infty}^{\infty} d\omega
e^{i\omega Q} Z(\beta,\omega).
\end{equation}
We can think of $\omega$ as a
canonical coordinate, in which case its canonically conjugate momentum
is $Q$, and we can think of (\ref{proj1}) as a Fourier transform.

Clearly, what we want to compare is the semi-classical approximation
to the partition functions $Z(\beta,Q)$ in the magnetic and the
electric case. For the magnetic case, the magnetic
Reissner-Nordstr\"om solution provides the saddle-point contribution
to the path integral, so
\begin{equation} \label{qpfun}
\ln Z(\beta,Q) = - I = - \beta M + {\cal A}_{bh}/4.
\end{equation}
In the electric case, the Fourier transform (\ref{proj1}) can also be
calculated by a saddle-point approximation. At the saddle-point,
$\omega = i \beta \Phi$, so
\begin{eqnarray}
\ln Z(\beta,Q) &=& - I + i \omega Q \nonumber \\
&=& - \beta (M-Q \Phi) + {\cal
A}_{bh}/4 + i \omega Q \\
&=&  - \beta M + {\cal A}_{bh}/4. \nonumber
\end{eqnarray}
Thus we see that the semi-classical approximation to the partition
function is the same for dual electric and magnetic black holes.

Alternatively, it is possible to construct a partition function
$Z(\beta,Q)$ for the electric case directly; that is, we can write
$Z(\beta,Q)$ in a path-integral form for a suitable choice of action
\cite{robb}. In the path integral, we want to use the action for which
it is natural to fix the boundary data on $\Sigma$ specified in the
path integral (\ref{pathin}). That is, we want to use an action whose
variation gives the Euclidean equations of motion when the variation
fixes these boundary data on $\Sigma$ \cite{brown}. If we consider the
action (\ref{act}), we can see that its variation will be
\begin{eqnarray}
\delta I &=& \mbox{ (terms giving the equations of motion) } \nonumber
\\ &&+ \mbox{
(gravitational boundary terms) } \nonumber \\
&& + \frac{1}{4\pi} \int_\Sigma d^3 x
\sqrt{h} F^{\mu\nu} n_\mu \delta A_\nu,
\end{eqnarray}
where $n_\mu$ is the normal to $\Sigma$ and $h_{ij}$ is the induced metric
on $\Sigma$ (see \cite{brown} for a more detailed discussion of the
gravitational boundary terms). Thus, the variation of (\ref{act})
will only give the equations of motion if the variation is at fixed
gauge potential on the boundary, $A_i$.

For the magnetic Reissner-Nordstr\"om solutions, fixing the gauge
potential fixes the charge on each of the black holes, as the magnetic
charge is just given by the integral of $F_{ij}$ over a two-sphere
lying in the boundary. However, in the electric case, fixing the gauge
potential $A_i$ can be regarded as fixing $\omega$.  Holding the
charge fixed in the electric case is equivalent to fixing $n_\mu
F^{\mu i}$ on the boundary, as the electric charge is given by the
integral of the dual of $F$ over a two-sphere lying in the
boundary. Therefore, the appropriate action is
\begin{equation} \label{elac2}
I_{el}  = I  - \frac{1}{4\pi} \int_{\Sigma} d^3 x \sqrt{h} F^{\mu\nu}
n_\mu A_\nu,
\end{equation}
as its variation is
\begin{eqnarray}
\delta I_{el} &=& \mbox{ (terms giving the equations of motion) }
\nonumber \\ &&+\mbox{
(gravitational boundary terms) } \nonumber \\
&&- \frac{1}{4\pi} \int_\Sigma d^3 x
\delta(\sqrt{h} F^{\mu\nu} n_\mu) A_\nu,
\end{eqnarray}
and so it gives the equations of motion when $\sqrt{h} n_\mu
F^{\mu i}$, and thus the electric charge, is held fixed. That is, if
we use (\ref{elac2}) in (\ref{pathin}) in the electric case, the
partition function we obtain is $Z(\beta,Q)$.

The observation that the magnetic charge must be imposed as a boundary
condition in the path integral has another, more troubling
consequence. In the derivation of the action for the asymptotically
flat black holes above, we have assumed that periodically-identified
flat space is a suitable background, so we can take $^2 K_0$ and $E_0$
in (\ref{ham}) to be the values of these quantities in flat
space. However, a suitable background is one which agrees with the
solution asymptotically; that is, it must satisfy the boundary
conditions in the path integral (\ref{pathin}). In the magnetic case,
periodically-identified flat space cannot satisfy these boundary
conditions, as it has no magnetic charge. Flat space is not a suitable
background to use in the evaluation of this action.  The best we can
do for single black holes is to compare the action of the non-extreme
black holes with the action of the extreme black hole of the same
charge, as this is a suitable background. It is natural to choose the
actions of the extreme black holes so that the partition functions for
fixed magnetic and electric charges are equal.  Such problems will not
arise in the case of pair creation in a cosmological background, as
the instantons are compact, so there is no need for a suitable
background solution to calculate the action.

\section{Reissner-Nordstr\"om de Sitter instantons}
\label{desitsec}

We will now describe the cosmological instanton, and
calculate its action. The Reissner-Nordstr\"om de Sitter metric
describes a pair of oppositely-charged black holes at antipodal points
in de Sitter space, as the Euclidean section has topology $S^2 \times
S^2$. The spatial sections therefore have topology $S^2 \times S^1$,
which may be thought of as a Wheeler wormhole, topology $S^2 \times
R^1$, attached to a spatial slice of de Sitter space, topology
$S^3$. The metric is
\begin{equation} \label{RNdesit}
ds^2 = -V(r) dt^2 + {dr^2 \over V(r)} + r^2 (d\theta^2 + \sin^2 \theta
d\phi^2),
\end{equation}
where
\begin{equation}
V(r) = 1 - {2M \over r} + {Q^2 \over r^2} - {\Lambda \over 3} r^2.
\end{equation}
We restrict consideration to just purely magnetically or purely
electrically charged solutions. The Maxwell field for the magnetically
charged solution is (\ref{mag}), and the Maxwell field for the
electrically charged solution is (\ref{elec}).  In general, $V(r)$ has
four roots, which we will label $r_1 <r_2 \leq r_3 \leq r_4$.  The two roots
$r_2$ and $r_3$ are the inner and outer black hole horizons, while
$r_4$ is the cosmological horizon. The smallest root $r_1$ is
negative, and thus has no physical significance.

We analytically continue $t \to i \tau$ to obtain a Euclidean
solution.  If the analytically continued metric is to be positive
definite, $r$ must lie between $r_3$ and $r_4$, where $V(r)$ is
positive. Then to have a regular solution, the surface gravities at
$r_3$ and $r_4$ must be equal, so that the potential conical
singularities at these two horizons can be eliminated with a single
choice of the period of $\tau$. This can be achieved in one of three
ways: either $r_3 = r_4$, $|Q| = M$, or $r_2 = r_3$
\cite{moss,romans,robb}.  Let us consider in detail the case where the
roots $r_3$ and $r_4$ are coincident, which is analogous to the
neutral black hole instanton studied in \cite{desit}. As in
\cite{desit}, the proper distance between $r=r_3$ and $r=r_4$ remains
finite in the limit $r_3 \to r_4$, as we can see by making a similar
change of coordinates. Let us set $r_3 = \rho - \epsilon$, $r_4 = \rho
+ \epsilon$. Then
\begin{equation}
V(r) = -\frac{\Lambda}{3r^2}(r - \rho - \epsilon)(r - \rho +
\epsilon)(r-r_1)(r-r_2).
\end{equation}
If we make a coordinate transformation
\begin{equation}
r = \rho + \epsilon  \cos \chi, \psi = A \epsilon \tau,
\end{equation}
where
\begin{equation}
A  =\frac{\Lambda}{3 \rho^2} (\rho-r_1)(\rho-r_2),
\end{equation}
then
\begin{equation}
V(r) \approx A \epsilon^2
\sin^2 \chi.
\end{equation}
Thus, in the limit $\epsilon \to 0$, the metric becomes
\begin{equation} \label{coinmetric}
ds^2  = {1 \over A} (d\chi^2 + \sin^2 \chi d\psi^2) + {1 \over B}
(d\theta^2 + \sin^2 \theta d\phi^2),
\end{equation}
where $\chi$ and $\theta$ both run from $0$ to $\pi$, and $\psi$ and
$\phi$ both have period $2\pi$.  This metric has been previously
mentioned in \cite{romans}. We assume that $B = 1/\rho^2 > A$ (this
corresponds to real $Q$, as we see below). The cosmological constant
is given by $\Lambda = (A+B)/2$, and the Maxwell field is
\begin{equation}
F = Q \sin \theta d \theta \wedge d \phi
\end{equation}
in the magnetically charged case, and
\begin{equation}
F = - iQ \frac{B}{A} \sin \chi d \chi \wedge d\psi
\end{equation}
in the electrically charged case, where $Q^2 = (B-A)/(2 B^2)$. This
metric is completely regular and, as the instanton is compact, it is
extremely easy to compute its action; it is
\begin{equation} \label{spaction}
I = -{1 \over 16 \pi} \int (R - 2 \Lambda -F^2)
  = -{ \Lambda V^{(4)} \over 8 \pi} \pm {Q^2 B^2 V^{(4)} \over 8 \pi},
\end{equation}
where $V^{(4)} =16 \pi^2/(AB)$ is the four-volume of the
instanton. The action for the magnetic case is thus $I = -2 \pi /B$,
and for the electric case the action is $I = -2 \pi /A$. Since the
action for the instanton describing the creation of neutral black
holes is $I = -2\pi/\Lambda$ \cite{desit}, we have $I_{magnetic} >
I_{neutral} > I_{electric}$. Further, $I_{de Sitter} > I_{electric}$
if $A < 2 \Lambda /3$. Since the action is supposed to give
the approximate rate for pair creation, this seems to say that de
Sitter space should be disastrously unstable to the pair creation of
large electrically charged black holes.

\section{Charge Projection for Reissner-Nordstr\"om de Sitter}
\label{cosm}

Clearly, there is an analogy between this problem and the
difficulty with the Reissner-Nordstr\"om solution, and so what
we need to do is to introduce a charge projection operator in the path
integral in the electric case. However, as the instanton is compact,
it looks like we don't have any boundary to specify boundary data on,
and in particular no notion of a chemical potential.

However, we are again forgetting something. The pair creation of black
holes in a de Sitter background is described, by the no-boundary
proposal, by the propagation from nothing to a three-surface
$\Sigma$ with topology $S^2 \times S^1$. This process is described
by a wavefunction
\begin{equation}
\Psi = \int d[g] d[A] e^{-I},
\end{equation}
where the integral is over all metrics and potentials on manifolds
with boundary $\Sigma$, which agree with the given boundary data on
$\Sigma$. This amplitude is dominated by a contribution from a
Euclidean solution which has boundary $\Sigma$ and satisfies the
boundary conditions there. For pair creation of black holes, the
instanton is in fact {\em half} of $S^2 \times S^2$.  In the
semi-classical approximation, $\Psi \approx e^{-I}$, where $I$ is the
action of this instanton.

In the usual approach reviewed in section \ref{chproj}, we take
advantage of the fact that the instanton is exactly half of the
bounce, so that the tunnelling rate is $\Psi^2 = Z= e^{-I_b}$, where
$I_b = 2I$ is the action of the bounce. This is helpful, as this
latter action is easier to calculate, but in passing from $\Psi$ to
$Z$ we have lost information about the boundary data on the surface on
which the bounce is sliced in half.  If there is a boundary at
infinity, this isn't very important,\footnote{It is easy to apply the
methods we outline below to re-derive the results of section
\ref{chproj} using the instanton (half the bounce) to describe
tunnelling from a spatial slice of hot flat space to a spatial slice
of electrically charged Reissner-Nordstr\"om.} but in the cosmological
case this information is crucial.

Consider the pair creation of charged black holes in a cosmological
background. Then $\Sigma$ has topology $S^2 \times S^1$, and the
boundary data on $\Sigma$ will be $h_{ij}$ and $A_i$, the three-metric
and gauge potential. In the magnetic case, we can again define the
charge by the integral of $F_{ij}$ over the $S^2$ factor (the charge
in this case is the magnitude of the charge on each of the black
holes), but in the electric case, we can
fix only the potential
\begin{equation} \label{omega}
\omega = \int A,
\end{equation}
where the integral is around the $S^1$ direction in $\Sigma$.  This
latter quantity is equal to the flux of the electric field across the
disk. Let $M_-$ be a Euclidean solution of the field equations which agrees
with the given data on $\Sigma$, which is its only boundary. If $M_-$
has topology $S^2 \times D^2$, which is the case we are interested in,
the $S^1$ direction in $\Sigma$ is the boundary of the two disk
$D^2$.

For the boundary data which describes a pair of charged black holes,
$M_-$ will just be half the $S^2 \times S^2$ Euclidean section of the
Reissner-Nordstr\"om de Sitter solution. Let us choose coordinates so
that the boundary $\Sigma$ corresponds to the surface $\psi=0, \psi
=\pi$ in the metric (\ref{coinmetric}), and so that the integral in
(\ref{omega}) is from the black hole horizon $\chi = \pi$ to the
cosmological horizon $\chi=0$ along $\psi=0$, and back along
$\psi=\pi$. The momentum canonically conjugate to $\omega$ is the
electric charge $Q$. Now
we are ready to make the Fourier transform
\begin{equation}
\Psi(Q,h_{ij}) = \frac{1}{2\pi} \int_{-\infty}^{\infty} e^{i\omega Q}
\Psi(\omega,h_{ij})
\end{equation}
to obtain the wavefunction in a definite charge sector in the electric
case.

We should make another Fourier transform, in both cases, as a natural
requirement on the three-surface $\Sigma$ is that its extrinsic
curvature vanish. This guarantees that $\Sigma$ bisects the bounce,
and ensures that our manifold can be matched smoothly onto a
Lorentzian extension.  We should therefore perform a Fourier transform
to trade $h_{ij}$ for its conjugate momentum $\pi^{ij} = \sqrt{h}(
K^{ij} - K h^{ij})$, where $K^{ij}$ is the extrinsic curvature of
$\Sigma$, and then set $\pi^{ij} =0$.  Thus
\begin{equation}
\Psi(Q,\pi^{ij}) = \frac{1}{2\pi}\int  d[h_{ij}] e^{i h_{ij}
\pi^{ij}} \Psi(Q,h_{ij}).
\end{equation}
In the saddle-point approximation,
\begin{equation}
\Psi(Q,\pi^{ij}=0) = \Psi(Q,h_{ij}=h_{ij}^0),
\end{equation}
where $h_{ij}^0$ is the induced metric on the three-surface $\psi=0,
\psi=\pi$ in the Reissner-Nordstr\"om de Sitter solution. That is,
because we are setting $\pi^{ij}=0$, there is no additional term in the
semi-classical value which arises from this transformation.

For the electrically charged Reissner-Nordstr\"om de Sitter instanton
(\ref{coinmetric}), the only vector potential which is regular
everywhere on $M_-$ is
\begin{equation} \label{dsvec}
A = i Q \frac{B}{A} \sin \chi \; \psi d\chi.
\end{equation}
We have to insist that the gauge potential be regular on the instanton
in the electric case to obtain gauge-independant results, as we can
only determine the gauge potential on the boundary.\footnote{This can
be clearly seen in the Reissner-Nordstr\"om case; we could set $A_t=0$
at infinity if we didn't insist that it be regular at the horizon.}
Note that there is {\it no} electric vector potential regular
everywhere on the Euclidean section of the electrically charged
Reissner-Nordstr\"om de Sitter solution, as (\ref{dsvec}) is not
periodic in $\tau$. Using (\ref{dsvec}), we see that in the
semi-classical approximation, $\omega = 2\pi i Q B/A$
and thus, in the electric case,
\begin{equation} \label{elres}
\ln \Psi(Q, \pi^{ij}=0) = -I +i \omega Q= \frac{\pi}{A} - \frac{2 \pi
Q^2 B}{A} = \frac{\pi}{B},
\end{equation}
as the action of $M_-$ is $-\pi/A$, half the action of the electric
instanton. For the magnetic solution,
\begin{equation} \label{magres}
\ln \Psi(Q, \pi^{ij}=0) = -I = \frac{\pi}{B},
\end{equation}
so the pair creation rate turns out to be identical in the two
cases. As $\Psi^2 \leq e^{2\pi/\Lambda} < e^{3\pi/\Lambda}$, these
processes are suppressed relative to both de Sitter space and the
neutral black hole instanton of \cite{desit}.

\section{Electric Ernst instantons}
\label{elecesec}

Black holes may be pair created by a background electromagnetic
field. An appropriate instanton which describes such pair creation is
provided by the Ernst solution, which represents a pair of
oppositely-charged black holes undergoing uniform acceleration in a
background electric or magnetic field. The magnetic case has been
extensively discussed, notably in \cite{dgkt,dggh,entar}. We now turn
to the consideration of the electric case, to see if the pair creation
rate is the same. An attempt was made to compare the electric case to
a charged star instanton in \cite{elec}. However, the action for Ernst
was not explicitly calculated. We find that the calculation of the
pair creation rate in this case introduces several new features, but
the pair creation rate given by $Z(\beta,Q)$ is identical to that
obtained in the magnetic case.

We will review the electric Ernst and Melvin solutions in this
section, and describe the calculation of the action in the following
section. The solution describing the background electric field is the
electric version of Melvin's solution \cite{melvin},
\begin{equation} \label{melvinm}
ds^2=\Lambda^2 \left(-dt^2+dz^2+d\rho^2\right)
+\Lambda^{-2}\rho^2 d\varphi^2,
\end{equation}
where
\begin{equation} \label{Llim}
\Lambda = 1+ \frac{\widehat{B}_M^2}{ 4} \rho^2 ,
\end{equation}
and the gauge field is
\begin{equation} \label{melving}
A_t = \widehat{B}_M z.
\end{equation}
The Maxwell field is $F^2 = -2\widehat{B}_M^2/\Lambda^4$, which is a
maximum on the axis $\rho=0$ and decreases to zero at infinity. The
parameter $\widehat{B}_M$ gives the value of the electric field on the
axis.

The metric for the electric Ernst solution is
\begin{eqnarray} \label{ernstm}
ds^2&=&(x-y)^{-2}A^{-2}\Lambda^2
\left[G(y)dt^2-G^{-1}(y)dy^2  \right. \\
&&+ \left. G^{-1}(x)dx^2\right]  +
(x-y)^{-2}A^{-2}\Lambda^{-2}G(x) d\varphi^2, \nonumber
\end{eqnarray}
where
\begin{equation}
G(\xi) = (1-\xi^2 - r_+ A \xi^3) (1+r_- A \xi),
\end{equation}
and
\begin{equation}
\Lambda=\left(1+\frac{1}{ 2}Bqx\right)^2+\frac{B^2}{ 4A^2(x-y)^2}G(x),
\end{equation}
while the gauge potential is \cite{elec}
\begin{eqnarray} \label{gpot}
A_t &=& -\frac{B G(y)}{2 A^2 (x-y)^2}\left[ 1 + \frac{1}{2} B q x +
\frac{1}{2} B q (x-y) \right] \\ &&- \frac{B}{2  A^2} (1+ r_+ A y) (1+r_-
Ay) \left( 1 - \frac{1}{2} B qy \right) + qy + k, \nonumber
\end{eqnarray}
where $k$ is a constant, and $q^2 = r_+r_-$.

If we label the roots of $G(\xi)$ by $\xi_1,\xi_2,\xi_3,\xi_4$ in
increasing order, then $x$ must be restricted to lie in $\xi_3 \leq x
\leq \xi_4$ to obtain a metric of the right signature. Because of the
conformal factor $(x-y)^{-2}$ in the metric, $y$ must be restricted to
$-\infty < y \leq x$. The axis $x=\xi_3$ points towards spatial
infinity, and the axis $x=\xi_4$ points towards the other black
hole. The surface $y=\xi_1$ is the inner black hole horizon, $y =
\xi_2$ is the black hole event horizon, and $y=\xi_3$ the acceleration
horizon. The black holes are non-extreme if $\xi_1 < \xi_2$, and
extreme if $\xi_1 = \xi_2$. Note that it is {\em not} possible to
choose $k$ so that $A_t$ vanishes at both $y=\xi_2$ and $y=\xi_3$. We
choose $k$ so that $A_t$ vanishes at $y=\xi_3$.

As discussed in \cite{dgkt}, to ensure that the metric is free of conical
singularities at both poles, $x=\xi_3, \xi_4$, we must impose the
condition
\begin{equation} \label{nonodes}
G^\prime(\xi_3)\Lambda(\xi_4)^2
= -G^\prime(\xi_4)\Lambda(\xi_3)^2,
\end{equation}
where $\Lambda(\xi_i)\equiv \Lambda(x=\xi_i)$. For later convenience,
we define $L \equiv \Lambda (x=\xi_3)$.  We also define a physical
electric field parameter $\widehat{B}_E = B G'(\xi_3) / 2
L^{3/2}$. When (\ref{nonodes}) is satisfied, the spheres are regular
as long as $\varphi$ has period
\begin{equation} \label{phiperiod}
\Delta\varphi=\frac{4\pi L^2}{
G^\prime(\xi_3) }\ .
\end{equation}

As in the magnetic case \cite{entar}, if we set $r_+ = r_- = 0$, the
Ernst metric reduces to the Melvin metric in accelerated form,
\begin{eqnarray}
ds^2 &=& \frac{\Lambda^2
}{ A^2 (x-y)^2} \left[ (1-y^2)
 dt^2 - \frac{dy^2}{  (1-y^2)}  \right. \\
&&+ \left. \frac{dx^2}{(1-x^2) }\right]
+ \frac{1-x^2 }{\Lambda^2  (x-y)^2 A^2}
d\varphi^2, \nonumber
\end{eqnarray}
where
\begin{equation}
\Lambda = 1+ \frac{\widehat{B}_E^2}{ 4} \frac{1-x^2}{ A^2 (x-y)^2}\ .
\end{equation}
The gauge field in this limit is
\begin{equation}
A_t = -\frac{\widehat{B}_E (1-y^2)}{2 A^2 (x-y)^2}.
\end{equation}
The acceleration parameter $A$ is now a coordinate degree of
freedom. Ernst also reduces to Melvin at large spatial distances, that
is, as $x,y \rightarrow \xi_3$.

We Euclideanize (\ref{ernstm}) by setting $\tau = it$. In the
non-extremal case, $\xi_1 < \xi_2$, the range of $y$ is taken to be
$\xi_2 \leq y \leq \xi_3$ to obtain a positive definite metric (we
assume $\xi_2 \ne \xi_3$).  To avoid conical singularities at the
acceleration and black hole horizons, we take the period of $\tau$ to
be
\begin{equation} \label{pert}
\beta = \Delta \tau = \frac{4 \pi }{ G'(\xi_3)}
\end{equation}
and require
\begin{equation} \label{nost}
G'(\xi_2)  = -G'(\xi_3),
\end{equation}
which gives
\begin{equation} \label{nostrtt}
 \xi_2 - \xi_1 = \xi_4 - \xi_3.
\end{equation}
The resulting Euclidean section has topology $S^2 \times S^2 -\{pt\}$, where
the point removed is $x=y=\xi_3$. This instanton is interpreted as
representing the pair creation of two oppositely charged black holes
connected by a wormhole.

If the black holes are extremal, $\xi_1=\xi_2$, the black hole event
horizon lies at infinite spatial distance from the acceleration
horizon, and gives no restriction on the period of $\tau$. The range
of $y$ is then $\xi_2 < y \leq \xi_3$, and the period of $\tau$ is
taken to be (\ref{pert}). The topology of the Euclidean section is $R^2
\times S^2 - \{ pt \}$, where the removed point is again
$x=y=\xi_3$. This instanton is interpreted as representing the pair
creation of two extremal black holes with infinitely long throats.

\section{Action in electric Ernst}
\label{elacsec}

Now, to calculate the pair creation rate, we need to calculate the
action for the instanton.  As in section \ref{cosm}, an instanton
describing the pair creation of black holes in a Melvin background is
given by cutting the Euclidean section above in half. That is, the
boundary $\Sigma$ that we want the instanton to interpolate inside of
consists of a three-boundary $S^3_\infty$ `at infinity', plus a
boundary $\Sigma_s$ which can be identified with the surface $\tau=0,
\tau=\beta/2$ in the Euclidean section.

Since we want to consider the pair creation rate at fixed electric
charge, the appropriate action is (\ref{elac2}).  That is, in the
instanton approximation, the partition function, and thus the pair
creation rate, is approximately given by $Z(\beta,Q) \approx
e^{-2I_{Ernst}}$, where $I_{Ernst}$ is the action (\ref{elac2}) of the
instanton.

Because the Euclidean section is not compact, the physical action is
only defined relative to a suitable background \cite{haho}, which in
this case is the electric Melvin solution. We need to ensure that we
use the same boundary $S^3_\infty$ in calculating the contributions to
the action from the Ernst and Melvin metrics. This is achieved by
insisting that the same boundary conditions are satisfied at the
boundary in these two metrics \cite{entar}.  That is, we insist that
the Ernst and Melvin solutions induce the same fields on the boundary
(up to contributions which vanish when we take the limit that the
boundary tends to infinity).

Let us take the boundary $S^3_\infty$ to lie at
\begin{equation}
x= \xi_3 + \epsilon_E \chi,\ \  y = \xi_3 + \epsilon_E (\chi -1),
\end{equation}
in the Ernst solution, and define new coordinates by
\begin{equation} \label{cchangei}
\varphi = \frac{ 2L^2 }{
G'(\xi_3)} \varphi',\ \tau = \frac{2 }{ G'(\xi_3)} \tau'.
\end{equation}
We assume that $S^3_\infty$ lies at
\begin{equation}
x = -1 + \epsilon_M \chi, y = -1 + \epsilon_M (\chi -1)
\end{equation}
in the accelerated coordinate system in the Melvin solution. The
metrics for the electric Ernst and Melvin solutions are the same as
for the magnetic solutions, so we know from \cite{entar} that the
induced metrics on the boundary can be matched by taking
\begin{equation} \label{abar}
\bar{A}^2 = -\frac{G'(\xi_3)^2 }{ 2 L^2 G''(\xi_3)} A^2,
\end{equation}
and
\begin{equation} \label{expans}
\epsilon_M = -\frac{G''(\xi_3) }{ G'(\xi_3)} \epsilon_E [1+
O(\epsilon_E^2)], \ \  \widehat{B}_M = \widehat{B}_E [1+ O(\epsilon_E^2)].
\end{equation}

However, we cannot match the gauge potentials at the same time. We
should work with a different gauge potential, as the gauge potential
(\ref{gpot}) is not regular at both the horizons in the spacetime. A
suitable gauge potential, which is regular everywhere on the instanton,
is
\begin{eqnarray} \label{gauge2}
A &=& -F_{x\tau} \tau dx - F_{y\tau} \tau dy \\ &=& i \tau \left[
\frac{B}{A^2(x-y)^3} G(y)  \left( 1 + \frac{1}{2} Bqx \right)\right]
dx \nonumber \\
&&+ i \tau \left[ q \left( 1 + \frac{1}{2} Bqx \right)^2 -
\frac{B}{A^2(x-y)^3} G(x)  \left( 1 + \frac{1}{2} Bqx \right) \right.
\nonumber \\
&& \left. + \frac{B}{2 A^2 (x-y)^2} G'(x)  \left( 1 + \frac{1}{2} Bqx \right) -
\frac{B^2 q}{4 A^2 (x-y)^2} G(x) \right] dy, \nonumber
\end{eqnarray}
and the
induced gauge potential on $S^3_\infty$ in the Ernst solution is
\begin{equation}
A_{\chi} = \frac{2i L^2 \tau' \widehat{B}_E}{A^2 \epsilon_E G'(\xi_3)}
\left[ 1+ \frac{G''(\xi_3)}{G'(\xi_3)} (\chi-1) \epsilon_E + \frac{Bq
 \chi \epsilon_E}{L^{1/2}}\right] ,
\end{equation}
while in the Melvin solution it is
\begin{equation}
A_\chi = \frac{i \tau \widehat{B}_M}{\bar{A}^2 \epsilon_M} \left[ 1 +
 \epsilon_M (\chi -1) \right],
\end{equation}
so they are {\em not} matched by (\ref{abar},\ref{expans}) (Note that,
even if we worked in the gauge (\ref{gpot}), the induced gauge
potentials on the boundary still wouldn't match). This seemed for a
long time to be an insuperable difficulty, but we have now realized
that, in the electric case, we no longer want to match $A_i$. Instead,
we should match $n_\mu F^{\mu i}$, and calculate the action
(\ref{elac2}), which will give the pair creation rate at fixed
electric charge.

The induced value of $n_\mu F^{\mu i}$ on $S^3_\infty$ in the
Ernst solution is
\begin{equation}
n_\mu F^{\mu t'} = \frac{ A \epsilon_E^{1/2} G'(\xi_3)^{1/2}
\widehat{B}_E}{2 L \lambda^3} \left[ 1 + \frac{G''(\xi_3)}{4
G'(\xi_3)} \epsilon_E (2\chi+1) \right],
\end{equation}
where
\begin{equation} \label{blambdaE}
 \lambda = \frac{\widehat{B}_E^2 L^2 }{ A^2 G'(\xi_3)
\epsilon_E} \chi +
\frac{\widehat{B}_E^2 L^2 G''(\xi_3) }{ 2 A^2 G'(\xi_3)^2} \chi^2 +1,
\end{equation}
while in the Melvin solution it is
\begin{equation}
n_\mu F^{\mu t} = \frac{\bar{A} \epsilon_M^{1/2}
\widehat{B}_M}{\sqrt{2} \Lambda^3} \left[ 1 - \frac{1}{4} \epsilon_M
(2\chi+1) \right],
\end{equation}
where
\begin{equation} \label{blambdaM}
\Lambda = \frac{\widehat{B}_M^2 }{ 2 \bar{A}^2 \epsilon_M} \chi -
\frac{\widehat{B}_M^2  }{ 4 \bar{A}^2} \chi^2 +1.
\end{equation}
We see that these two quantities are indeed matched by
(\ref{abar},\ref{expans}).

The action (\ref{elac2}) of the region of the Ernst solution inside
$\Sigma$ can be written as a surface term, as we can see by writing it
in covariant form:
\begin{eqnarray}
I_{el} &=& \frac{1}{16 \pi} \int d^4 x \sqrt{g}(-R + F^2) -
\frac{1}{8\pi} \int_{\Sigma} d^3 x \sqrt{h}  K \\ &&- \frac{1}{4\pi}
\int_{\Sigma} d^3 x \sqrt{h} F^{\mu\nu} n_\mu A_\nu \nonumber
\\ &=& - \frac{1}{8\pi}\int_{\Sigma} d^3 x \sqrt{h}  K - \frac{1}{8\pi}
\int_{\Sigma} d^3 x \sqrt{h} F^{\mu\nu} n_\mu A_\nu, \nonumber
\end{eqnarray}
as the volume integral of $R$ is zero by the field equations, and the
volume integral of the Maxwell Lagrangian $F^2$ can be converted to a
surface term by the field equations. The explicit surface term in
(\ref{elac2}) just reverses the sign of the electromagnetic surface
term obtained from the $F^2$ volume integral; that is, it has the
effect of reversing the sign of the electromagnetic contribution to
the action.

Using the gauge choice (\ref{gauge2}), we see that the action is
\begin{eqnarray}
I_{el} &=& - \frac{1}{8\pi}\int_{S^3_\infty} d^3 x \sqrt{h} K -
\frac{1}{8\pi} \int_{\Sigma_s} d^3 x \sqrt{h} F^{\mu\nu} n_\mu A_\nu,
\\ &=& - \frac{1}{8\pi}\int_{S^3_\infty} d^3 x \sqrt{h} K -
\frac{1}{16\pi} \frac{\beta}{2} \int dx dy d\varphi \sqrt{g} F^2
 \nonumber
\end{eqnarray}
In the first line, we have used the fact that the extrinsic curvature
of $\Sigma_s$ vanishes, and that $n^\mu A^\nu F_{\mu\nu} = 0$ on
$S^3_\infty$; in the second line, we used (\ref{gauge2}). This is the
same as the expression for the action in \cite{entar} (as the Maxwell
term changes sign), and the matching conditions are the same, so we
can use the calculation of the action in \cite{entar} to conclude that
\begin{equation}
I_{Ernst} =  \frac{\pi L^2}{A^2 G'(\xi_3) (\xi_3-\xi_1)}.
\end{equation}
The pair creation rate is approximately $e^{-2I_{Ernst}}$, so it is
thus identical to that for the magnetic case. Note that this applies
to both extreme and non-extreme black holes. In particular, the pair
creation of non-extreme black holes is enhanced over that of extreme
black holes by a factor of $e^{{\cal A}_{bh}/4}$, as it was in the
magnetic case \cite{entar}.

\section{Entropy of charged black holes}
\label{entsec}

We turn now to a discussion of the thermodynamics of black
holes. Consider first the asymptotically-flat black holes. In the
electric case, one can calculate the partition function for the grand
canonical ensemble at temperature $T$ and electrostatic chemical
potential $\Phi$. One does a path integral over all fields that have
given period and potential at infinity. In the semi-classical
approximation, the dominant contribution to the path integral comes
from solutions of the field equations with the given boundary
conditions. These are the electrically charged Reissner-Nordstr\"om
solutions. The semi-classical approximation to the partition function
is $Z(\beta,\omega) \approx e^{- I}$, where $I$ is the action of the
solution. But in the grand canonical ensemble, $\ln Z(\beta,\omega)
=- \Omega/ T$, where $\Omega$ is a thermodynamic potential \cite{actionint},
\begin{equation}
\Omega = M - TS - Q\Phi.
\end{equation}
Comparing this with the expression (\ref{elac}) for the action, one
finds that the $M$ and $Q \Phi$ terms cancel, leaving the entropy
equal to a quarter of the area, $S = {\cal A}_{bh}/4$, as expected.

In the case of magnetic black holes, the entropy still comes out to be
${\cal A}_{bh}/4$, but the calculation is rather different.  Since the
magnetic charge is defined by the asymptotic form of the potential, or
equivalently, by the choice of the electromagnetic fiber bundle, there
is a separate canonical ensemble for each value of the magnetic
charge, which is necessarily an integer, unlike the electric charge of
an ensemble, which is a continuous variable.\footnote{We might add
that the angular momentum of a black hole is a continuous variable,
and is not quantised, because it is the expectation value in an
ensemble, not a quantum number of a pure state. In the grand canonical
ensemble, one therefore has to introduce angular velocity $\Omega $ as
a chemical potential for angular momentum, like one introduces the
electrostatic potential as a chemical potential for charge. The
Hamiltonian gets an additional $\Omega J$ surface term.  } In the
magnetic case, the charge is always quantised, even for an
ensemble. There is thus no need for a chemical potential for magnetic
charge. That is, the partition function $Z$ depends on the charge, and
should therefore be interpreted as the canonical partition function,
so $\ln Z(\beta,Q) = -F/T$, where $F$ is the free energy,
\begin{equation}
F = M -TS,
\end{equation}
while the action is (\ref{magac}), and the entropy is therefore again
$S ={\cal A}_{bh}/4$.

We should note that, if we make the Fourier transform (\ref{proj1}) in
the electric case, the partition function $Z(\beta,Q)$ is also
interpreted as the canonical partition function. Therefore, this
Fourier transform may be thought of as a Legendre transform giving the
free energy in terms of the thermodynamic potential $\Omega$
\cite{quhair},
\begin{equation}
F(\beta,Q) =\Omega(\beta,\Phi) + Q \Phi.
\end{equation}
The result we get for the entropy doesn't depend on whether we work
with the partition function $Z(\beta,\omega)$ or $Z(\beta,Q)$ in the
electric case.

The absence of the Maxwell surface term in the Hamiltonian for
magnetic black holes means that they have higher action than their
electric counter parts. For an extreme black hole, the region swept
out by the surfaces of constant time covers the whole instanton, so
$I=\beta H$ \cite{haho,entar}. Now, as before, $H=M$ for a magnetic
black hole, so the action of an extreme magnetic black hole is $I=\beta
M$, where $\beta$ is now an {\em arbitrary} period with which one can
identify an extreme black hole. For the electric case, $H= M-Q\Phi$,
but $Q=M$ implies $r_+=M$, and thus $\Phi=1$, so $I=0$ for an extreme
electric black hole. Both of these actions are proportional to
$\beta$. This means that if you substitute the actions into the usual
formula
\begin{equation} \label{canent}
S = - \left( \beta \frac{\partial}{\partial \beta} -1 \right) \ln Z
\end{equation}
for the entropy, where $Z \approx e^{-I}$, you find that both extreme
electric and magnetic black holes have zero entropy, as previously
announced \cite{entar}.

Now for the cosmological black holes, we again need to work with the
wavefunction $\Psi(Q,\pi^{ij}=0)$ rather than the partition function
$Z$. Because it does not depend on the temperature, $\Psi^2$ can be
interpreted as the microcanonical partition function, or density of
states \cite{brown}. In fact, it should be clear that $\Psi$
represents a closed system, and the partition function should just be
interpreted as counting the number of states, so the entropy should be
$S = \ln Z$, or more accurately,
\begin{equation}
S = 2 \ln \Psi(Q,\pi^{ij}=0).
\end{equation}
Note that it is $\Psi(Q,\pi^{ij}=0)$, and not $\Psi(\omega,\pi^{ij}=0)$,
which gives the microcanonical partition function. If we evaluate this
entropy in the semi-classical approximation, where $\Psi(Q,\pi^{ij}=0)$
is given by (\ref{elres},\ref{magres}), we get
\begin{equation}
S = 2\pi /B =  {\cal A}/4
\end{equation}
in both cases, as there are two horizons, which both have area
$4\pi/B$, so the total area ${\cal A} = 8\pi/B$. That is, we find that
the usual relation between entropy and area holds here too, despite
the fact that $\Psi$ has a very different interpretation in this case.

\section{Discussion}

We have seen that the action of dual electric and magnetic solutions
of the Einstein-Maxwell equations differ. This is presumably a general
property of $S$-duality, and initially led us to wonder whether
$S$-duality could be a symmetry of a quantum theory given by a path
integral. However, we found that despite the difference in actions,
the semi-classical approximation to the partition function in a
definite charge sector was the same for dual electric and magnetic
solutions. In particular, we found that the rate at which electrically
and magnetically charged black holes are created in a background
electromagnetic field or in a cosmological background is the same.

The pair creation of both types of charged black holes in a
cosmological background by the instanton studied here is suppressed
relative to de Sitter space, as we might expect, and it is also more
strongly suppressed than the creation of neutral black holes. The
instantons describing pair creation of black holes in a cosmological
background are studied in more detail in \cite{robb}. For all the
instantons, the rate at which the pair creation occurs is suppressed
relative to de Sitter space.

These calculations are all just semi-classical, but they do seem to
offer some encouragement to the suggestion that electric-magnetic
duality is more than just a symmetry of the equations of motion.  The
conclusion seems to be that duality is a symmetry of the quantum
theory, but in a very non obvious way. As Einstein said, God is
subtle, but he is not malicious.

\acknowledgements

The authors were greatly helped by discussions with John Preskill
while they were visiting Cal Tech.  S.F.R. thanks the Association of
Commonwealth Universities and the Natural Sciences and Engineering
Research Council of Canada for financial support.

\end{document}